\def\b{\bibitem}
\def\boldphi{\mbox{\boldmath $\phi$}}
\def\boldvarphi{\mbox{\boldmath $\varphi$}}
\documentstyle[aps,prb,eqsecnum,epsf,psfig,floats]{revtex}
\begin{document}
\def\SNG{{\em Physical Review Style and Notation Guide}}
\def\LUG {{\em \LaTeX{} User's Guide \& Reference Manual}}
\def\btt#1{{\tt$\backslash$\string#1}}%
\def\REVTeX{REV\TeX}
\def\AmS{{\protect\the\textfont2
        A\kern-.1667em\lower.5ex\hbox{M}\kern-.125emS}}
\def\AmSLaTeX{\AmS-\LaTeX}
\def\BibTeX{\rm B{\sc ib}\TeX}
\twocolumn[\hsize\textwidth\columnwidth\hsize\csname@twocolumnfalse%
\endcsname
\title{On the critical behavior of disordered quantum magnets: The relevance
       of rare regions}
\author{Rajesh Narayanan}
\address{Department of Physics and Materials Science Institute,
                                      University of Oregon, Eugene, OR 97403}
\author{Thomas Vojta} 
\address{Department of Physics and Materials Science Institute,
                                      University of Oregon, Eugene, OR 97403\\
 and Institut f{\"u}r Physik, TU Chemnitz, D-09107 Chemnitz, FRG}
\author{D. Belitz}
\address{Department of Physics and Materials Science Institute,
                                      University of Oregon, Eugene, OR 97403}
\author{T.R. Kirkpatrick}
\address{Institute for Physical Science and Technology, and Department of 
         Physics, \\
         University of Maryland, College Park, MD 20742}
\date{\today}
\maketitle
\begin{abstract}
The effects of quenched disorder on the critical properties of itinerant 
quantum antiferromagnets and ferromagnets are considered. Particular attention 
is paid to locally ordered spatial regions that are formed in the presence of
quenched disorder even when the bulk system is still in the paramagnetic phase.
These rare regions or local moments are reflected in the existence of spatially
inhomogeneous saddle points of the Landau-Ginzburg-Wilson functional. We derive
an effective theory that takes into account small fluctuations around {\em all}
of these saddle points. The resulting free energy functional 
contains a new term in addition to those obtained within the conventional 
perturbative approach, and it comprises what would be considered 
non-perturbative effects within the latter.
A renormalization group analysis shows that in the case 
of antiferromagnets, the previously found critical fixed point is unstable with
respect to this new term, and that no stable critical fixed point exists at 
one-loop order. This is contrasted with the case of itinerant ferromagnets, 
where we find that the previously found critical behavior is unaffected by the 
rare regions due to an effective long-ranged interaction between the order 
parameter fluctuations.
%
%
\end{abstract}
]
\section{Introduction}
\label{sec:I}

The influence of static or quenched disorder on the critical properties
of a system near a continuous phase transition is a very interesting
problem in statistical mechanics. While it was initially suspected that
quenched disorder always destroys any critical point,\cite{Grinstein}
this was soon found to not necessarily be the case. Harris\cite{Harris}
found a convenient criterion for the stability of a given critical
behavior with respect to quenched disorder: If the correlation length
exponent $\nu$ obeys the inequality $\nu\geq 2/D$, with $D$ the spatial
dimensionality of the system, then the critical behavior is unaffected
by the disorder. In the opposite case, $\nu < 2/D$, the disorder modifies
the critical behavior.\cite{controversy}
This modification may either (i) lead to a new critical
point that has a correlation length exponent $\nu \geq 2/D$ and is thus
stable, or (ii) lead to an unconventional critical point where the
usual classification in terms of power-law critical exponents looses 
its meaning, or (iii) lead to the destruction of a sharp phase transition. 
The first possibility is realized in the conventional theory of random-$T_c$
classical ferromagnets,\cite{Grinstein} and the second one is probably
realized in classical ferromagnets in a random
field.\cite{random-field} The third one has occasionally been attributed to 
the exactly solved McCoy-Wu model.\cite{McCoyWu} This is misleading,
however, as has recently been emphasized in Ref.\ \onlinecite{DSF};
there is a sharp, albeit unorthodox, transition in that
model, and it thus belongs to category (ii). 

Independent of the question of if and how the critical behavior is affected,
disorder leads to very interesting phenomena as a phase transition is
approached. Disorder in general decreases the critical temperature $T_c$ from 
its clean value $T_c^0$. In the temperature region $T_c<T<T_c^0$ the system
does not display global order, but in an infinite system one will find 
arbitrarily large regions that are devoid of impurities, and hence show 
local order, with a small but non-zero probability that usually decreases
exponentially with the size of the region. These static disorder fluctuations
are known as `rare regions', and the order parameter fluctuations induced
by them as `local moments' or `instantons'. Since they are weakly coupled,
and flipping them requires to change the order parameter in a whole
region, the local moments have very slow dynamics. 
Griffiths\cite{Griffiths} was the first to show that they lead to a
non-analytic free energy everywhere in the region $T_c<T<T_c^0$, which
is known as the Griffiths phase, or, more appropriately, the Griffiths
region. In generic classical systems this is a weak effect, since the
singularity in the free energy is only an essential one. An important
exception is the McCoy-Wu model,\cite{McCoyWu} which is a $2$-$D$
Ising model with bonds that are random along one direction, but identical
along the second direction. The resulting infinite-range correlation of
the disorder in one direction leads to very strong effects.
As the temperature is lowered through the Griffiths
region, the local moments cause the divergence of an increasing number of 
higher order susceptibilities, 
$\chi^{(n)} = \partial^n M/\partial B^n$ ($n\geq 2$), with $M$ the order
parameter and $B$ the field conjugate to it, starting with large $n$. 
Even the average susceptibility proper, $\chi^{(1)}$, diverges at a 
temperature $T_{\chi} > T_c$, although the average order parameter does 
not become non-zero until the temperature
reaches $T_c$. This is caused by rare fluctuations in the susceptibility
distribution, which dominate the average susceptibility and make it very
different from the typical or most probable one.

Surprisingly little is known about the influence of the Griffiths region
and related phenomena on the critical behavior. Recent work\cite{Dotsenko} 
on a random-$T_c$ classical Ising model has suggested that it can be profound, 
even in this simple model where the conventional theory predicts standard 
power-law critical behavior, albeit with critical exponents that are 
different from the clean case. The authors of
Ref.\ \onlinecite{Dotsenko} have shown that the conventional theory is
unstable with respect to perturbations that break the replica symmetry.
By approximately taking into account the rare regions, which are
neglected in the conventional theory, they found a new term in the
action that actually induces such perturbations. In some systems
replica symmetry breaking is believed to be associated with activated,
i.e. non-power law, critical behavior. Reference \onlinecite{Dotsenko}
thus raised the interesting possibility that, as a result of rare-region
effects, the random-$T_c$ classical Ising model shows activated critical 
behavior, as is believed to be the case for the random-field classical
Ising model,\cite{random-field} although in the case of the random-$T_c$ 
model no final conclusion about the fate of the transition could be 
reached.

Griffiths regions also occur in the case of quantum phase transitions 
(QPTs), i.e. transitions that occur at zero temperature as a function of 
some non-thermal control parameter.\cite{QPT,CN} Their consequences for
the critical behavior are even less well investigated than in the
classical case, with the remarkable exception of certain
$1$-$D$ systems. Fisher\cite{DSF} has
investigated quantum Ising spin chains in a transverse random field.
These systems are closely related to the classical McCoy-Wu model, with
time in the quantum case playing the role of the `ordered direction' in
the latter. He has found activated critical
behavior due to rare regions. This has been confirmed by numerical
simulations.\cite{APY} Other recent simulations\cite{APY2d} suggest that this
type of behavior may not be restricted to $1$-$D$ systems, raising
the possibility that exotic critical behavior dominated by rare regions
may be generic in quenched disordered quantum systems, independent of the
dimensionality and possibly also of the type of disorder.

In this paper we consider this problem analytically for two QPTs in
$D>1$. We first concentrate on a simple model for a quantum antiferromagnet.
Previous work,\cite{afm} which did not take into account rare regions,
had found a transition with some surprising properties. One of our goals
is to check whether these results survive taking into account rare regions.
We find that they do not; the previously found critical fixed point is
unstable with respect to the rare regions, and one finds runaway flow
in all of physically accessible parameter space. We will discuss possible 
interpretations of this 
result. We then show that the critical behavior of itinerant quantum 
{\em ferro}magnets is {\em not} affected by the rare regions, in sharp 
contrast to the antiferromagnetic case. A brief report of some of our 
results has been given previously in Ref.\ \onlinecite{us_Letter}.

The paper is organized as follows. In Sec.\ \ref{sec:II} we derive an
effective action for an itinerant antiferromagnet in the presence of
rare regions. In Sec.\ \ref{sec:III} we perform a one-loop renormalization
group analysis of this action, and show that there is no stable critical
fixed point to that order. In Sec.\ \ref{sec:IV} we perform an analogous
analysis for itinerant ferromagnets and show that the previously found
critical fixed point is stable with respect to the rare region effects.
In Sec.\ \ref{sec:V} we discuss our results. Various technical points
are relegated to three appendices.

\section{An effective action for disordered antiferromagnets}
\label{sec:II}

\subsection{The model}
\label{subsec:II.A}

Our starting point is Hertz's action\cite{Hertz} for an itinerant quantum
antiferromagnet. It is a $\phi^4$-theory with a $p\,$-component order 
parameter field $\boldphi$ whose expectation value is proportional to the
staggered magnetization. The bare two-point vertex function is
\begin{equation}
\Gamma_0({\bf q},\omega_n) = t_0 + {\bf q}^2 + \vert\omega_n\vert\quad,
\label{eq:2.1}
\end{equation}
with $t_0$ the mean distance from the mean-field critical point. ${\bf q}$ is
the wavevector, and $\omega_n$ denotes a bosonic Matsubara frequency. We
measure ${\bf q}$ and $\omega_n$ in suitable microscopic units to make
them dimensionless. As
in Ref.\ \onlinecite{afm}, we modify this action by adding disorder in the
form of a `random-mass' or `random-temperature' term. That is, we add to
$t_0$ a random function of position, $\delta t({\bf x})$, which obeys a
distribution with zero mean and variance $\Delta$,
\begin{mathletters}
\label{eqs:2.2}
\begin{equation}
\langle\delta t({\bf x})\rangle = 0\quad,
\label{eq:2.2a}
\end{equation}
\begin{equation}
\langle\delta t({\bf x})\,\delta t({\bf y})\rangle = \Delta\,\delta({\bf x}
                                                        - {\bf y})\quad.
\label{eq:2.2b}
\end{equation}
\end{mathletters}%
For the sake of simplicity, we have taken the distribution to be
delta-correlated. The two-point vertex now reads
\begin{eqnarray}
\Gamma({\bf x}-{\bf y},\tau-\tau')&=&\Gamma_0({\bf x}-{\bf y},\tau - \tau')
\nonumber\\
  &&+ \delta({\bf x}-{\bf y})\,\delta(\tau - \tau')\,\delta t({\bf x})\quad.
\label{eq:2.3}
\end{eqnarray}
Here $\tau$ denotes imaginary time, and $\Gamma_0({\bf x},\tau)$ is the
Fourier transform of $\Gamma_0({\bf q},\omega_n)$ in Eq.\ (\ref{eq:2.1}).
The action reads
\begin{mathletters}
\label{eqs:2.4}
\begin{equation}
S[\boldphi ] = S_{\rm G}[\boldphi ]
   +  u\int dx\,\left(
   \boldphi (x)\cdot\boldphi (x)\right)^2\quad,
\label{eq:2.4a}
\end{equation}
with the Gaussian part given by
\begin{equation}
S_{\rm G}[\boldphi ] = \frac{1}{2} \int dx\,dy\ 
  \boldphi (x)\,\Gamma(x-y)\,\boldphi (y)\quad.
\label{eq:2.4b}
\end{equation}
\end{mathletters}%
Here we have introduced a four-vector notation, $x\equiv ({\bf x},\tau)$,
$\int dx \equiv \int d{\bf x}\int_0^{1/T} d\tau$, and
we use units such that $\hbar = k_{\rm B} = 1$.

At this point, the conventional procedure would be to integrate
out the quenched disorder by means of the replica trick.\cite{Grinstein}
This would lead to an effective action that does not contain the disorder 
explicitly any longer, and that therefore does not easily allow for 
saddle-point solutions that are not spatially homogeneous. While
the effective action would still be exact, this latter property
would make it hard to incorporate the physics we are concentrating
on in this paper. We will therefore take a different approach, and consider
saddle-point solutions of the model, Eqs.\ (\ref{eqs:2.4}), {\em before} 
integrating out the disorder. Our procedure roughly follows the one by
Dotsenko et al.\cite{Dotsenko} for classical magnets. As we will see,
however, there are important differences between the classical and
quantum cases.

\subsection{Saddle-point solutions}
\label{subsec:II.B}

Let us consider saddle-point solutions of Eqs.\ (\ref{eqs:2.4}) that are
time independent. For simplicity, we also consider a scalar field, $p=1$,
that we denote by $\phi ({\bf x})$. It will be obvious how to generalize
the following considerations to the case $p>1$. With these restrictions, 
the saddle-point equation reads
\begin{equation}
\left(t_0 + \delta t(\bf x) - \partial_{\bf x}^2\right)\,\phi_{\rm sp}({\bf x})
     + 4\,u\,\phi_{\rm sp}^3({\bf x}) = 0\quad.
\label{eq:2.5}
\end{equation}
Although $\phi_{\rm sp}^{(1)}({\bf x}) \equiv 0$ is of course always a
solution, inhomogeneous
solutions also exist provided that $\delta t({\bf x})$ has `troughs' that are
sufficiently wide and deep.\cite{TroughFootnote} In Appendix\ \ref{app:A}
we demonstrate this for a one-dimensional toy problem. We have solved
Eq.\ (\ref{eq:2.5}) numerically for rotationally invariant potentials
$\delta t({\bf x}) = f(\vert{\bf x}\vert)$, and have found behavior
that is qualitatively the same as in the one-dimensional model. Thus,
if $\delta t({\bf x})$ has one sufficiently deep and wide trough, there
will be a solution of Eq.\ (\ref{eq:2.5}) that is exponentially small
everywhere except within the trough region, where it shows a single hump.
There are actually two equivalent single-hump solutions, one positive, and
the other negative. We denote the positive one by
$\phi_{\rm sp}^{(2)}({\bf x}) \equiv \psi^{(+)}({\bf x})$. As is intuitively
obvious, and
demonstrated in Appendix\ \ref{app:A}, it leads to a lower free energy
than the homogeneous solution $\phi_{\rm sp}^{(1)}({\bf x}) \equiv 0$.
Although in principle any saddle point can be used as the
starting point for a loop expansion,
it is therefore reasonable to assume that the inhomogeneous one will already
incorporate physics that would be much harder to obtain if we expanded about
the homogeneous saddle point.

Next consider a potential $\delta t({\bf x})$ that contains many troughs
that support an essentially non-zero local order parameter field. This will
result in a saddle-point solution that contains many regions of local order,
which we will refer to as `islands'. Of course, for an arbitrary potential
$\delta t({\bf x})$ it is not possible to solve Eq.\ (\ref{eq:2.5}) in
closed form. However, as long as the concentration of the islands is low,
as will always be the case sufficiently deep in the disordered phase (i.e.,
for sufficiently large $t_0$), the values of $\phi_{\rm sp}$ outside
of the islands will still be exponentially small. If $\delta t$ has
troughs leading to $N$ islands, which individually would result in positive
saddle-point solutions $\psi_i^{(+)}({\bf x})$, $(i=1,\ldots,N)$,
it is therefore a reasonable
approximation to write the global saddle-point solution as a linear
superposition of the $\psi_i^{(+)}$.
For independent islands, there are
actually $2^N$ equivalent saddle points, which we write as
\begin{eqnarray}
\phi^{(a)}_{\rm sp}({\bf x}) \equiv \Phi^{(a)}({\bf x})
  &=&\sum_{i=1}^{N} \sigma_i^{a}\,\psi_i^{(+)}({\bf x})
\nonumber\\
  &\equiv& \sum_{i=1}^{N} \psi_i({\bf x})\quad,
\label{eq:2.6}
\end{eqnarray}
where $a = 1,\ldots, 2^N$ numbers the equivalent saddle points,
and the $\sigma_i^{a}$ are random numbers whose values are either
$+1$ or $-1$. They thus obey a probability distribution
\begin{mathletters}
\label{eqs:2.7}
\begin{equation}
P[\{\sigma_i^{a}\}] = \prod_i \pi(\sigma_i^{a})\quad,
\label{eq:2.7a}
\end{equation}
with
\begin{equation}
\pi(\sigma) = \frac{1}{2}\,
   \left[\delta(\sigma - 1) + \delta(\sigma + 1)\right] \quad.
\label{eq:2.7b}
\end{equation}
Alternatively, one can consider the $\psi_i({\bf x})$ random functions
that are equal to either plus or minus $\psi_i^{(+)}({\bf x})$.

For later reference, let us briefly discuss the effects of a weak interaction
between the islands as a result of the exponentially small overlap between
the functions $\psi_i({\bf x})$ centered on different islands. One effect will
be that the total amount of the order parameter on each island will not
necessarily be equal to plus or minus the amount resulting from that island
being ideally ordered, but that small deviations from this amount will be
possible. If we still assume that the islands are statistically independent,
we can model this effect by using a probability distribution for the
$\sigma_i$ that is given by Eq.\ (\ref{eq:2.7a}) with a distribution 
$p(\sigma)$ that is a broadended version of the bimodal delta-distribution 
$\pi(\sigma)$ given in Eq.\ (\ref{eq:2.7b}). 
We thus generalize Eqs.\ (\ref{eq:2.7a},\ref{eq:2.7b}) to
\begin{equation}
P[\{\sigma_i^{a}\}] = \prod_i p(\sigma_i^{a})\quad.
\label{eq:2.7c}
\end{equation}
\end{mathletters}%
For our purposes we will not need to specify the distribution $p(\sigma)$
explicitly. It will turn out that an interaction between the
islands, no matter how small, leads to new physics compared to a model
where these interactions are neglected.

We also note that the islands will have some dynamics, both due to
interactions between the islands and due to interactions between an
island and its immediate neighborhood. In principle, one could try to
build this effect into the saddle-point approximation by looking for
time dependent saddle points. However, this dynamics is expected to
be very slow due to the inertia of the islands. Moreover, the zero
frequency component in a frequency expansion is expected to yield the
dominant effect. We therefore restrict ourselves to static saddle points.

\subsection{The partition function for a given disorder realization}
\label{subsec:II.C}

Of the $2^N$ saddle-point solutions $\Phi^{(a)}({\bf x})$ discussed
in the previous subsection, let us pick one, say $\Phi^{(1)}$, to expand about:
\begin{mathletters}
\label{eqs:2.8}
\begin{equation}
\phi (x) = \Phi^{(1)}({\bf x}) + \varphi(x)\quad.
\label{eq:2.8a}
\end{equation}
Then the partition function can be written
\begin{equation}
Z[\delta t({\bf x})] = \int D[\varphi(x)]\,e^{-S[\Phi^{(1)}({\bf x}) +
   \varphi(x),\delta t({\bf x})]}\quad,
\label{eq:2.8b}
\end{equation}
\end{mathletters}%
where we show $\delta t({\bf x})$ explicitly as an argument to emphasize that
we are still working with a fixed realization of the disorder.
Equation (\ref{eq:2.8b}) is exact as long as the integral extends over
{\em all} fluctuations $\varphi(x)$ of the field configuration.
However, in practice the integral over $\varphi(x)$ cannot be
performed exactly, and in a perturbative treatment one restricts oneself
to small deviations $\varphi(x)$ from the chosen saddle point. Typical
pairs of saddle points picked from the $2^N$ $\Phi^{(a)}$ represent field
configurations that are globally very different. They will thus be
far apart in configuration space, with large energy barriers between them.
(We will justify this statement in Sec.\ \ref{subsubsec:V.A.2} below.)
Expanding about one of the saddle points, as in Eqs.\ (\ref{eqs:2.8}),
is therefore not expected to yield a good representation of the partition
function if one evaluates the functional integral in Eq.\ (\ref{eq:2.8b})
perturbatively. On the other hand, the same argument suggests that we can
simply sum the contributions to $Z$ obtained by expanding about {\em all}
of the $2^N$ saddle points, provided that we restrict ourselves to
small fluctuations about each saddle point,
\begin{equation}
Z[\delta t({\bf x})] \approx \sum_{a=1}^{2^N} \int_{<} D[\varphi(x)]
  \ e^{-S[\Phi^{(a)}({\bf x}) + \varphi(x),\delta t({\bf x})]}\quad.
\label{eq:2.9}
\end{equation}
Here $\int_{<}$ indicates an integration over small fluctuations only.
Apart from a normalization factor, this procedure amounts to an
arithmetic average over the perturbative contributions coming from the
vicinities of all saddle points. This average is our approximative way
of taking into account non-perturbative effects.

Since we are interested in the effects of fluctuations about the saddle
points, we next subtract the saddle point action from the exponent in
Eq.\ (\ref{eq:2.9}).\cite{SubtractionFootnote} That is, we write
\begin{equation}
Z[\delta t({\bf x})] \approx \sum_{a}\int_{<}
   D[\varphi(x)]\
   e^{-\Delta S[\Phi^{(a)}({\bf x}),\varphi(x),\delta t({\bf x})]},
\label{eq:2.10}
\end{equation}
where
\begin{eqnarray}
\Delta S[\Phi^{(a)},\varphi,\delta t]&\equiv&
   S[\Phi^{(a)} + \varphi,\delta t] - S[\Phi^{(a)},\delta t]
\nonumber\\
&=&S[\varphi(x)] + 4u\int dx\,\varphi^3(x)\,
     \Phi^{(a)} ({\bf x})
\nonumber\\
    &&+  6u\int dx\,\varphi^2(x)\,
        \Phi^{(a)}({\bf x})^2\quad.
\label{eq:2.11}
\end{eqnarray}

So far we have implicitly assumed that there is no interaction between the 
islands. In reality, there will be a small interaction, one effect of which
will be to replace the bimodal distribution, 
Eqs.\ (\ref{eq:2.7a},\ref{eq:2.7b}), by the broadended distribution given
in Eq.\ (\ref{eq:2.7c}). The sum over $a$ in Eq.\ (\ref{eq:2.10})
is then replaced by an integral over the $\sigma_i^{a}$, weighted
by the distribution $p(\sigma)$. The partition function can now be
written as
\begin{mathletters}
\label{eqs:2.12}
\begin{equation}
Z[\delta t({\bf x})] \approx \int D[\varphi(x)]\ e^{-(S[\varphi(x)]
   + \delta S[\varphi(x)])}\quad,
\label{eq:2.12a}
\end{equation}
with the correction to the action, $\delta S$, given by
\begin{eqnarray}
e^{-\delta S[\varphi(x)]}&=&\int\prod_{i=1}^{N} d\sigma_i\
   \prod_j p(\sigma_j)\qquad\qquad\qquad\qquad\qquad
\nonumber\\
&\times& e^{-4u\int dx\,\varphi^3(x)
         \,\sum_i\sigma_i\psi_i^{(+)}({\bf x})}
\nonumber\\
&\times& e^{-6 u\,\int dx\,\varphi^2 (x)\,
   \sum_{ij}\sigma_i\sigma_j\psi_i^{(+)}({\bf x})\psi_j^{(+)}({\bf x})
                  }\ .
\nonumber\\
\label{eq:2.12b}
\end{eqnarray}

As mentioned previously, it is crucial to incorporate a small interaction 
between the islands. Indeed, if we used the distribution function, 
Eqs.\ (\ref{eqs:2.7}), for the
$\sigma_i$ that is appropriate for non-interacting islands, then we
could do the $\sigma$-integral in Eq.\ (\ref{eq:2.12b}) exactly.
As we will show in Sec.\ \ref{sec:III}, the resulting action
would not lead to new physical effects compared to Ref.\ \onlinecite{afm}.
Apart from the broadening of the distribution, there
are other, similar effects of the island-island interaction that
we will neglect. For instance, in the second exponent on the
right-hand-side of Eq.\ (\ref{eq:2.12b}), the absence of any overlap
between $\psi_i^{(+)}$ and $\psi_j^{(+)}$ for $i\neq j$ makes the spatial 
integral in that term vanishes unless $i=j$. This is no longer true
for interacting islands. However, we will neglect this effect and still
take this term to be proportional to $\delta_{ij}$. Equation (\ref{eq:2.12b})
can then be written
\begin{eqnarray}
\delta S[\varphi(x)]&=&-\sum_i \ln \left\langle
  e^{-4u\int dx\,\varphi^3(x)\,\psi_i^{(+)}({\bf x})\,\sigma_i}\right.
\nonumber\\
 &&\times\left. e^{-6u\int dx\,\varphi^2(x)\,
    \left(\psi_i^{(+)}({\bf x})\right)^2 \sigma_i^2} \right\rangle\quad.
\label{eq:2.12c}
\end{eqnarray}
\end{mathletters}%
Here $\langle\ldots\rangle$ denotes an average over the $\sigma$ with
respect to the broadened bimodal distribution $p(\sigma)$.
The $\sigma$-average is carried out by means of a cumulant
expansion in powers of our order parameter
fluctuations $\varphi(x)$. To the order $\varphi^4$ we obtain
\begin{eqnarray}
\delta S[\varphi(x)]&=& 6u\int dx\,\varphi^2(x)\sum_i
     \left(\psi_i^{(+)}({\bf x})\right)^2\,\langle\sigma_i^2\rangle
\nonumber\\
&&+4u\int dx\,\varphi^3(x)\sum_i \psi_i^{(+)}({\bf x})\,\langle\sigma_i\rangle
\nonumber\\
&&-18u^2\int dx\,dy\,\varphi^2(x)\,\varphi^2(y)\,
\nonumber\\
&&{\hskip -9mm}\times\sum_i
  \left(\psi_i^{(+)}({\bf x})\right)^2\,\left(\psi_i^{(+)}({\bf y})\right)^2\,
  \left(\langle\sigma_i^4\rangle - \langle\sigma_i^2\rangle^2\right)\quad.
\nonumber\\
\label{eq:2.13}
\end{eqnarray}
Now $\langle\sigma_i\rangle=0$, $\langle\sigma_i^2\rangle > 0$ and
$\langle\sigma_i^4\rangle - \langle\sigma_i^2\rangle^2 \equiv c_i > 0$.
The last relation is only valid for a broadened distribution
$p(\sigma)$ which arises from interactions between the islands.
For the original distribution $\pi(\sigma)$,
$c_i = 0$ and so the $O(\varphi^4)$ term would vanish.
If we collect all contributions to the action for one particular
disorder configuration, we obtain
\begin{eqnarray}
&S[\varphi& (x)] + \delta S[\varphi (x)]
\nonumber\\
 &=&\frac{1}{2} \int dx dy\ \varphi(x)\,\Gamma(x-y)\,\varphi(y)
       + u \int dx\ \varphi^4(x) 
\nonumber\\
 &+& 18 u^2 \int dx dy\ \varphi^2(x)\,\varphi^2(y) \sum_i c_i\,
  \left(\psi_i^{(+)}({\bf x}) \psi_i^{(+)}({\bf y})\right)^2\,.
\nonumber\\ 
\label{eq:2.14}
\end{eqnarray}
Here we have used the fact that the first term in (\ref{eq:2.13}) only
renormalizes the random-mass term in the Gaussian action.
We will show in Sec.\ \ref{sec:III} that truncating the
action at $O(\varphi^4)$ is justified since
all higher order terms are irrelevant (in a power-counting sense) with
respect to both the Gaussian fixed point and
the antiferromagnetic fixed point found in Ref.\ \onlinecite{afm}.

\subsection{The effective action}
\label{subsec:II.D}

So far we have considered one particular realization of the disorder. In 
order to derive an effective action we now need to perform the disorder 
average. It is important to remember that the Landau functional,
Eq.\ (\ref{eq:2.14}), depends on the disorder in two different ways:
explicitly through the random mass in the Gaussian action, and
implicitly through the saddle-point solutions $\psi_i^{+}({\bf x})$
that depend on $\delta t({\bf x})$.

The quenched disorder average over $\delta t(x)$ is performed via
the replica trick,\cite{Grinstein} which is based on the identity
\begin{equation}
\{ \log\, Z\, \}_{\delta t} = \lim_{n \rightarrow 0} {\{ Z^n\, \}_{\delta t}
- 1 \over n}\quad.
\label{eq:2.15}
\end{equation}
Here $\{ \ldots \}_{\delta t}$ denotes the disorder average.
This results in an effective action $S_{\rm eff}$ which is
defined by
\begin{eqnarray}
\{ Z^n \}_{\delta t} &=&  \int \prod_{\alpha=1}^n
   D[\varphi^{\alpha}(x)]\left\{ e^{ -\sum_{\alpha} ( S[\varphi^{\alpha}(x)]
   +\delta S[\varphi^{\alpha}(x)] ) }\right\}_{\delta t}\nonumber\\
&\equiv&\int \prod_{\alpha=1}^n
   D[\varphi^{\alpha}(x)] ~ e^{-S_{\rm eff}[\varphi^{\alpha}(x)]}\quad.
\label{eq:2.16}
\end{eqnarray}

In the absence of $\delta S$, carrying out the disorder average yields
the usual terms that are familiar from the conventional theory. Up to
$O(\varphi^4)$ they are:
\begin{eqnarray}
&~&\frac 1 2 \sum_\alpha \int dx\, dy\ 
    \varphi^\alpha(x)\, \Gamma_0(x-y)\, \varphi^\alpha(y) \nonumber\\
&+& u \sum_\alpha \int dx (\varphi^\alpha(x))^4
    \nonumber\\
&-& \Delta  \sum_{\alpha,\beta} \int d{\bf x}\, d\tau\, d\tau'\ 
    ((\varphi^\alpha({\bf x},\tau))^2\, (\varphi^\beta({\bf x},\tau'))^2\ .
\label{eq:2.17}
\end{eqnarray}
Taking into account the additional term, $\delta S[\varphi^{\alpha}(x)]$, 
is more subtle since the functions $\psi_i (x)$ are implicit functions of 
$\delta t(x)$. We handle this problem by means of a cumulant expansion.
To lowest order, the contribution of $\delta S$ to the effective action
is just the disorder average of $\delta S$,
\begin{mathletters}
\label{eqs:2.17}
\begin{equation}
\{\delta S\}_{\delta t} = w \int dx dy\ \varphi^2(x)\,\varphi^2(y)\,
              D_{\rm isl}^{(2)}({\bf x,y})\quad,
\label{eq:2.18a}
\end{equation}
where $w \propto u^2$, and
the correlation function 
\begin{equation}
D_{\rm isl}^{(2)}({\bf x,y}) = \left\{\sum_i c_i\,\left(
    \psi_i^{(+)}({\bf x})\,\psi_i^{(+)}({\bf y})\right)^2\right\}_{\delta t}
\label{eq:2.18b}
\end{equation}
\end{mathletters}%
essentially describes the probability
for ${\bf x}$ and ${\bf y}$ to belong to the same island. The properties of 
these correlation functions depend on the precise nature of the disorder.
If the microscopic disorder $\delta t({\bf x})$ is short-range
correlated, as we have assumed in our model, then the island size
distribution will generically fall off exponentially for large
sizes. In this case the correlation function
$D_{\rm isl}^{(2)}$ is also short-ranged in space. Keeping only the
leading term in a gradient expansion, we can then replace it
by a spatial $\delta$-function. The case of an island size distribution 
that has a power-law tail (e.g. due to long-range correlations in the 
microscopic disorder) is discussed in Appendix \ref{app:B}.

Collecting all contributions to the effective action $S_{\rm eff}$ up to 
$O(\varphi^4)$, absorbing a constant into $w$, and restoring the
vector nature of the order parameter field, we finally obtain
\begin{eqnarray}
S_{\rm eff}[\boldvarphi^\alpha(x)] &=&
  {1\over 2} \sum_\alpha\,  \int dx\, dy\,\Gamma_0(x-y)\,
  \boldvarphi^\alpha(x)\cdot \boldvarphi^\alpha(y)\
\nonumber\\
&&\hskip -22pt + u\sum_\alpha \int d{\bf x}\, d\tau
\left(\boldvarphi^\alpha({\bf x},\tau)\cdot
   \boldvarphi^\alpha({\bf x},\tau)\right)^2\
\nonumber\\
&&\hskip -22pt -\sum_{\alpha,\beta} (\Delta + w\,\delta_{\alpha\beta})
   \int d{\bf x}\,d\tau\,d\tau^{\prime}
\nonumber\\
&&\hskip -12pt\times(\boldvarphi^\alpha ({\bf x},\tau)\cdot
     \boldvarphi^\alpha ({\bf x},\tau))\,
   (\boldvarphi^\beta ({\bf x},\tau^{\prime})
      \cdot\boldvarphi^\beta ({\bf x},\tau^{\prime}))\,.
\nonumber\\
\label{eq:2.19}
\end{eqnarray}
The $w$-term is generated by taking into account the inhomogeneous saddle
points. A perturbative expansion about the homogeneous saddle point, as
was performed in Ref.\ \onlinecite{afm}, misses this term.
It has the time structure of the random-mass or $\Delta$-term, and the 
replica structure of the quantum fluctuation or $u$-term. We will further
discuss its physical meaning in Sec.\ \ref{subsec:V.A}. 
In the following section 
we will show that the critical fixed point found in Ref. \onlinecite{afm}
is unstable with respect to this new term in the action.

\section{Renormalization group analysis}
\label{sec:III}

\subsection{Tree-level analysis}
\label{subsec:III.A}

Let us first justify our truncation of the Landau expansion in 
Sec.\ \ref{sec:II} by showing that all terms of higher than quartic order
in $\boldvarphi$ are irrelevant (in the renormalization group sense) by
power counting with respect to the critical fixed point of 
Ref.\ \onlinecite{afm}. To this end, we analyze the effective action,
$S_{\rm eff}$, Eq.\ (\ref{eq:2.19}), at tree level.

Let us denote the scale dimension of any quantity $Q$ by $[Q]$, and define 
the scale dimension of  a length $L$ to be $[L] = -1$. The scale dimension 
of the imaginary time is $[\tau] = -z$, which defines the dynamical critical 
exponent $z$. We first analyze the Gaussian fixed point. From the structure
of the two-point vertex function $\Gamma_0$ given in Eq.\ (\ref{eq:2.1}), 
we see that $\omega_n$ scales like $q^2$. This implies $z = 2$. The scale 
dimension of the field $\boldvarphi$ is found (from the requirement that
the action must be dimensionless) to be $[\boldvarphi(x)] = D/2$.
The scale dimensions of the coefficients of the terms of $O(\boldvarphi^4)$ 
in Eq.\ (\ref{eq:2.19}) are found to be $[u] = 2-D$, and 
$[\Delta] = [w] = 4-D$. Thus, $u$ is irrelevant with respect to the Gaussian 
fixed point as long as $D>2$, while $\Delta$ and $w$ are relevant for $D<4$. 
The Gaussian fixed point is therefore unstable, and we will have to perform 
a loop expansion close to $D=4$ in the next subsection.

We now show that all terms of $O(\boldvarphi^6)$ and higher are irrelevant 
with respect to the Gaussian fixed point. First of all, there are the 
conventional terms of the schematic form 
\begin{equation}
u_{2m} \int dx\,\varphi^{2m}(x)\quad,
\label{eq:3.1}
\end{equation}
with coupling constants $u_{2m}$ ($u_4\equiv u$). These are irrelevant since 
$[u_{2m}] = 2 - (m-1)D<0$. In addition to these terms, the cumulant expansion 
of (\ref{eq:2.12c}) generates higher order terms with more time integrations 
than the conventional terms for a given power of $\varphi$. For instance, at
$O(\varphi^6)$ we have two terms,
\begin{mathletters}
\label{eqs:3.2}
\begin{equation}
8u^2 \int dx\,dy\ \varphi^3(x)\,\varphi^3(y) \sum_i \psi_i^{(+)}({\bf x})\,
   \psi_i^{(+)}({\bf y})\,\langle\sigma_i^2\rangle \quad,
\label{eq:3.2a}
\end{equation}
and
\begin{eqnarray}
&-& 36u^3 \int dx\,dy\,dz\ \varphi^2(x)\,\varphi^2(y)\,\varphi^2(z) \sum_i
   \left(\psi_i^{(+)}({\bf x})\right)^2\,
\nonumber\\
 &&\qquad \times \left(\psi_i^{(+)}({\bf y})\right)^2\,
   \left(\psi_i^{(+)}({\bf z})\right)^2\,
\nonumber\\
 &&\qquad\quad\times
   \left(\langle\sigma_i^6\rangle - 
         3\langle\sigma_i^4\rangle\langle\sigma_i^2\rangle
               + 6\langle\sigma_i^2\rangle^3\right)\quad.
\label{eq:3.2b}
\end{eqnarray}
\end{mathletters}%
Upon averaging over the disorder these terms become
\begin{mathletters}
\label{eqs:3.3}
\begin{equation}
v_{6}\,\int dx_1\,dx_2\,\varphi^3(x_1)\,\varphi^3(x_2)\,
C_{\rm isl}^{(2)}({\bf x}_1,{\bf x}_2)\quad,
\label{eq:3.3a}
\end{equation}
and
\begin{eqnarray}
&w_{6}&\,\int dx_1\,dx_2\,dx_3\,\varphi^2(x_1)\,\varphi^2(x_2)\,\varphi^2(x_3)
\nonumber\\
&&\qquad\qquad\qquad\times D_{\rm isl}^{(3)}({\bf x}_1,{\bf x}_2,{\bf x}_3)
                                                          \quad,
\label{eq:3.3b}
\end{eqnarray}
\end{mathletters}%
respectively, with $v_6 \propto u^2$, $w_6 \propto u^3$.
The correlation functions
$C_{\rm isl}^{(2)}({\bf x},{\bf y})$
and $D_{\rm isl}^{(3)}({\bf x},{\bf y}, {\bf z})$
are defined analogously to $D_{\rm isl}^{(2)}({\bf x,y})$ in
Eq.\ (\ref{eq:2.18b}), and are related to the probability
for ${\bf x},{\bf y}$ and ${\bf x},{\bf y},{\bf z}$, respectively, to belong 
to the same island.
We again concentrate on the generic case where the island size
distribution falls of exponentially for large islands
(for a discussion of other cases, see Appendix \ref{app:B}).
In this case both correlation functions are 
short-ranged and can be localized for power-counting purposes. 
This effectively leaves only one spatial integral in Eqs.\ (\ref{eq:3.3a}) 
and (\ref{eq:3.3b}). Therefore, the scale dimensions of the coefficients are
$[v_{6}] = 2(2-D)$ and $[w_{6}] = 2(3-D)$. Consequently, both terms
are irrelevant with respect to the Gaussian fixed point near $D=4$.

More generally, we obtain from Eq.\ (\ref{eq:2.12c}) terms that contain
powers of $\varphi^3$, terms that contain powers of $\varphi^2$, and
mixed terms that contain both $\varphi^3$ and $\varphi^2$. For power
counting purposes, the most relevant term for a fixed power of $\varphi$
is the one with the most time integrations. For even powers of $\varphi$,
these are the terms
\begin{equation}
w_{2m} \int dx_1\,\ldots dx_m\,\varphi^2(x_1)\ldots\varphi^2(x_m)\,
   D^{(m)}_{\rm isl}({\bf x}_1,\ldots,{\bf x}_m).
\label{eq:3.4}
\end{equation}
Localizing the correlation function $D^{(m)}$ we find for the scale
dimension of the coupling constant $[w_{2m}] = 2m - (m-1)D$. Terms with
odd powers of $\varphi$ are always less relevant than the preceding term
of even order. We conclude that all terms of higher than quartic order
are irrelevant with respect to the Gaussian fixed point near $D=4$.

So far we have determined the scale dimensions with respect to the Gaussian 
fixed point. At the non-trivial critical fixed point discussed in 
Ref. \onlinecite{afm} the anomalous dimension of the field $\varphi$ is
$\eta = 0 + O(\epsilon^2)$ since, as in the ordinary $\phi^4$ theory, 
there is no wavefunction renormalization at 1-loop order.
This implies that all results on the irrelevancy of the terms
of order $\varphi^6$ and higher carry over from the Gaussian
fixed point to the non-trivial critical fixed point
found in Ref. \onlinecite{afm}.

\subsection{Perturbation theory, and flow equations}
\label{subsec:III.B}

In the last subsection we have shown that the Gaussian fixed point
is unstable for $D<4$. We must therefore carry out a loop expansion for
the effective action, Eq.\ (\ref{eq:2.19}). To control the perturbation 
theory we consider $D = 4 - \epsilon$ spatial dimension and $\epsilon_{\tau}$ 
time dimensions.\cite{Cardy} This leads to the
replacement of $\int d\tau$ by $\int d\tau\, \tau^{\epsilon_\tau - 1}$.
At the Gaussian fixed point the scale dimension of the field $\varphi$ is
now $[\varphi] = (d + z\epsilon_{\tau} - 2)/2$. In the same vein the scale
dimensions of $u$, $\Delta$, and $w$ are $[u] = \epsilon - z\epsilon_{\tau}$,
$[\Delta] = \epsilon$, and $[w] = \epsilon$, respectively. The perturbation
theory becomes a double expansion in $\epsilon$ and $\epsilon_{\tau}$.

\begin{figure}
\epsfxsize=7.0cm
\centerline{\epsffile{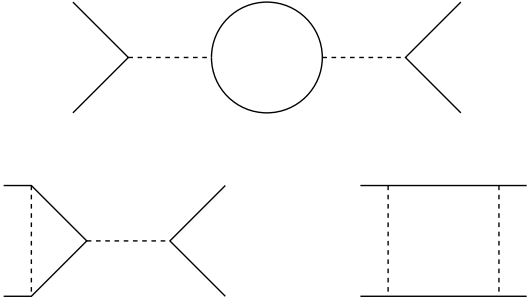}}
\vskip 10mm
\caption{The three diagram structures that contribute to the flow equations
           shown in Eqs.\ (\protect\ref{eqs:3.5}). The dashed lines stand for
           any of the vertices whose coupling constants are $u$, $\Delta$,
           or $w$.}
  \label{fig:1}
\end{figure}

To obtain the renormalization group flow equations, we perform a
frequency-momentum shell RG procedure.\cite{Hertz}
The diagrams that contribute to the renormalization of the coupling
constants $u$, $\Delta$, and $w$ are shown in Fig.\ \ref{fig:1}.
To one-loop order, we obtain the following flow equations,
\begin{mathletters}
\label{eqs:3.5}
\begin{equation}
\frac{du}{dl} = (\epsilon - 2\epsilon_{\tau})\,u
                              - 4\,(p+8)\,u^2 + 48\,u\,\Delta \quad,
\label{eq:3.5a}
\end{equation}
\begin{equation}
\frac{d\Delta}{dl} = \epsilon\Delta + 32\,\Delta^2 - 8\,(p+2)\,u\,\Delta
+ \frac{4p}{T^{\epsilon_{\tau}}}\, (2\,\Delta - w)\, w\ ,
\label{eq:3.5b}
\end{equation}
\begin{eqnarray}
\frac{dw}{dl} = \epsilon\,w + \frac{4p}{T^{\epsilon_{\tau}}}\,w^{2} -
                              8\,(p+2)\,u\,w
           &+&24\, (2\, \Delta\,- w)\, w\
\nonumber\\
           &+& 8\, w^2\quad.
\label{eq:3.5c}
\end{eqnarray}
\end{mathletters}%
The mass $t$ of the two-point vertex, which describes the distance from the 
critical point, is of course also renormalized. However, since we are 
interested in the stability of a critical fixed point, it suffices to
consider the flow on the critical surface. The factors of 
$T^{-\epsilon_{\tau}}$ in Eqs.\ (\ref{eqs:3.5}) arise from the fact that 
some diagrams that contain the $w$-vertex lead to Matsubara frequency sums 
without an accompanying temperature factor. Since the critical surface for
the quantum phase transition is defined by $T=0$ in addition to $t=0$,
the natural coupling constant for the $T=0$ flow is
$\bar{w} = w\,T^{-\epsilon_{\tau}}$. Putting 
$T=0$,\cite{ClassicalLimitFootnote} the flow equations
can then be rewritten in the form
\begin{mathletters}
\label{eqs:3.6}
\begin{equation}
\frac{du}{dl} = (\epsilon - 2\epsilon_{\tau})\,u
                              - 4\,(p+8)\,u^2 + 48\,u\,\Delta \quad,
\label{eq:3.6a}
\end{equation}
\begin{equation}
\frac{d\Delta}{dl} = \epsilon\,\Delta + 32\,\Delta^2 -
           8\,(p+2)\,u\,\Delta + 8\, p\, \Delta\,\bar{w}\quad,
\label{eq:3.6b}
\end{equation}
\begin{equation}
\frac{d\bar{w}}{dl} = (\epsilon - 2\epsilon_{\tau})\,\bar{w}
      + 4\,p\,\bar{w}^{2} - 8\,(p+2)\,u\,\bar{w} + 48\,\Delta
       \,\bar{w}\ .
\label{eq:3.6c}
\end{equation}
\end{mathletters}%

\subsection{Fixed points and their stability}
\label{subsec:III.C}

The flow equations, Eqs.\ (\ref{eqs:3.6}), possess eight fixed points.
The fixed-point values of the coupling constants, and the corresponding 
eigenvalues of the linearized RG transformation are listed in 
Table \ref{tab:1}.
\begin{table*}[t]
\renewcommand{\arraystretch}{2.0}
\begin{tabular}{c|ccc|ccc}
\multicolumn{1}{c|}{FP No.\hskip 3mm} 
      & \multicolumn{3}{c|}{FP values$\hskip 7mm$} 
      & \multicolumn{3}{c}{eigenvalues}                                   \\
      & $u^*$ 
      & $\Delta^*$ 
      & $w^*\hskip 9mm$ 
      & $\lambda_1$ 
      & $\lambda_2$ 
      & $\lambda_w$                                                       \\
\hline
1     & 0 
      & 0 
      & 0$\hskip 9mm$ 
      & $\epsilon - 2\epsilon_{\tau}$ 
      & $\epsilon$ 
      & $\epsilon - 2\epsilon_{\tau}$                                     \\
2     & $\frac{\epsilon - 2\epsilon_{\tau}}{4(p+8)}$
      & $0$
      & $0 \hskip 9mm$ 
      & $-(\epsilon - 2\epsilon_{\tau})$ 
      & $\frac{8(p+4)\epsilon_{\tau} - (p-4)\epsilon}{p+8}$
      & $\frac{(p-4)(\epsilon - 2\epsilon_{\tau})}{p+8}$       \\
3     & $0$
      & $-\epsilon/32$
      & $0 \hskip 9mm$
      & $2\epsilon_{\tau} - \epsilon/2$
      & $-\epsilon$
      & $2\epsilon_{\tau} - \epsilon/2$                      \\
4     & $\frac{(\epsilon + \epsilon_\tau)}{16\,(p - 1)}$
      & $\frac{(4 - p)\epsilon + 4(p +2)\epsilon_\tau}{64\,(p - 1)}$
      & $0 \hskip 9mm$
      & $\frac{-A + \sqrt {A^2 - B}}{p - 1}$
      & $\frac{-A - \sqrt {A^2 - B}}{p - 1}$
      & $\frac{(4 - p)\epsilon +
                4(4 - p)\epsilon_{\tau})}{4\,(p - 1)}$          \\
\hline
5     & $0$
      & $0$
      & $-(\epsilon - 2\epsilon_{\tau})/4p \hskip 9mm$
      & $\epsilon/4 - \epsilon_{\tau}/2$
      & $4 \epsilon_{\tau} - \epsilon$
      & $- (\epsilon - 2\epsilon_{\tau})$                       \\
6     & $\frac{\epsilon - 2\epsilon_{\tau}}{4\,(p + 8)}$
      & $0$
      & $\frac{(p - 4)\, (\epsilon - 2\epsilon_{\tau})}{4p\,(p + 8)}
                                                         \hskip 9mm$
      & $- (\epsilon - 2\epsilon_{\tau})$
      & $\frac{2\, (p + 4 )\,2\,\epsilon_{\tau} - (p - 4)\epsilon}{p + 8}$
      & $\frac{(p - 4)\, (\epsilon - 2\, \epsilon_{\tau})}{p + 8}$  \\
7     & $0$
      & $(4\epsilon_{\tau} - \epsilon)/64$
      & $-(4\epsilon_{\tau} + \epsilon)/16p\hskip 9mm$
      & $\epsilon_{\tau} + \epsilon/4$
      & $- \epsilon$
      & $2\epsilon_{\tau} - \epsilon/2$          \\
8     & $\frac{\epsilon + 4\epsilon_{\tau}}{8\,(10 - p)}$
      & $\frac{((p - 4)\epsilon + 24\epsilon_{\tau})}{32\,(10 -p)}$
      & $\frac{(p - 4)(\epsilon + 4\epsilon_{\tau})}{8p\,(10 - p)}\hskip 9mm$
      & $\frac{-C + \sqrt {C^2 - D}}{p - 1}$
      & $\frac{-C - \sqrt {C^2 - D}}{2\,(10- p)} $
      & $\frac{(p - 4)(\epsilon + 4\epsilon_{\tau})}{2\,(10 - p)}$
\end{tabular}
\vspace{2mm}
\caption{Fixed points of the flow equations, Eqs. (\ref{eqs:3.6}) and
 the eigenvalues of the corresponding linearized RG transformation.
 $p$ is the number of order parameter components. $\lambda_{1,2}$ are
 the eigenvalues in the absence of $w$, and $\lambda_w$ is the additional
 eigenvalue. $A$, $B$, $C$, and $D$ are defined as 
 $A = (3\epsilon - 4\epsilon_{\tau})p +16\epsilon_{\tau}$,
 $B = 16(p-1)\,(\epsilon + 4\epsilon_{\tau})\,
                         [(4-p)\epsilon + 4(p+2)\epsilon_{\tau}]$,
   $ C = (16-p)\epsilon + 4(p-4)\epsilon_{\tau}$,
   and $D = (10 - p)\, (p + 8)\, (\epsilon + 4\epsilon_{\tau})\,
   [(p - 4)\epsilon + 24)\epsilon_{\tau}]$.
}
\label{tab:1}
\end{table*}
Four of the fixed points (Nos. 1--4 in Table \ref{tab:1})
have a zero fixed-point value of ${\bar w}$, $\bar{w}^*=0$.
These are the fixed points studied before
in Ref. \onlinecite{afm}. The other four fixed points have
$\bar{w}^* \neq 0$.

Let us first consider fixed point No. 4. This is the critical fixed point
that was found within the conventional approach.\cite{afm} We find that
the local moments, represented by the $w$-term, render this fixed point 
unstable for $p<4$, since in this case the third eigenvalue, $\lambda_{w}$, 
is positive. However, for $p>4$ the $w$-term is irrelevant with respect to
this fixed point, and the fixed point is stable for $4<p<p_c$. To one-loop
order, and for $\epsilon = \epsilon_{\tau}$, $p_c=16$.\cite{afm}

A stability analysis for the new fixed points shows that they are 
all unstable for $p<4$ with the exception of No. 8. At this fixed point, 
$w$ has a negative value, while Eq.\ (\ref{eq:2.13}), with reasonable 
assumptions about the distribution $p(\sigma)$,
yields a positive value for the bare value of $w$.
Since the structure of the flow equations does not allow $w$ to change
sign, we conclude that this fixed point is unphysical for generic realizations
of the disorder. It is interesting to note, however, that fixed point No. 8
is stable against replica symmetry breaking (see Appendix \ref{app:C}).

For $p<4$, and to one-loop order in our double expansion in powers of
$\epsilon$ and $\epsilon_{\tau}$, there is thus no stable fixed point.
Consistent with this, a numerical solution of the flow equations, 
Eqs.\ (\ref{eqs:3.6}), shows runaway flow in all of physical parameter
space. We will discuss the physical meaning of this result in 
Sec.\ \ref{sec:V} below.

\section{The case of itinerant ferromagnets}
\label{sec:IV}

In Ref. \onlinecite{fm} a generalized LGW functional for the ferromagnetic 
transition in a disordered itinerant electron system was derived starting 
from a fermionic description. The effects of rare regions were not
explicitly considered in this work. Here we show that although the rare
regions were neglected in the explicity calculations in that paper,
the effective field theory derived in Ref. \onlinecite{fm} still
contains these effects. We will further show that taking them into account
does not change the previous conclusions. 

We first briefly recall the effective action that was derived in
Ref.\ \onlinecite{fm}. In the long-wavelength and low-frequency limit, 
the replicated action is given by
\begin{eqnarray}
S_{\rm eff,1} &=&\frac{1}{2}\sum_\alpha \int dx_1\,dx_2\ \Gamma_0(x_1-x_2)\,
   {\mathbf M}^\alpha (x_1)\cdot {\mathbf M}^\alpha (x_2)
\nonumber\\
&&+\sum_\alpha \int
dx_1\,dx_2\,dx_3\,dx_4\ u_4(x_1,x_2,x_3,x_4)
\nonumber\\
&&\hskip 20pt \times ({\mathbf M}^{\alpha}(x_1)\cdot{\mathbf M}^{\alpha}(x_2))\ 
({\mathbf M}^{\alpha}(x_3)\cdot{\mathbf M}^{\alpha}(x_4))
\nonumber\\
&&-\Delta \sum_{\alpha,\beta }\int d{\mathbf x}\,d\tau\,d\tau^{\prime}\ 
   ({\mathbf M}^{\alpha}({\mathbf x},\tau))^2\,({\mathbf M}^{\beta}
      ({\mathbf x},\tau^{\prime}))^2\quad.
\nonumber\\
\label{eq:4.1}
\end{eqnarray}
Here ${\mathbf M}$ is the order parameter field whose expectation value is
the magnetization, and $u$ and $\Delta$ are coupling constants that in 
general are wavenumber and frequency dependent. An important point is that 
these coupling constants in general do not exist in the limit of zero
frequencies and wavenumbers, i.e. the effective action describes a non-local
field theory. This is because in the process of deriving a LGW functional
that depends only on the order parameter field,
soft (viz., diffusive) fermionic degrees of freedom have been integrated out.
In writing Eq.\ (\ref{eq:4.1}) we have used that 
the coupling constant $\Delta $ is finite in the long-wavelength 
limit, so that it can be treated as a number. $u_4$, on the other hand, 
is singular in this limit, see Eq.\ (\ref{eq:4.3}) below. For small 
wavenumbers the Fourier transform of
the two-point vertex $\Gamma_0$ is given by
\begin{mathletters}
\label{eqs:4.2}
\begin{equation}
\Gamma_0({\bf q},\omega_n) = t_0 + u_2({\bf q}) + 
   \vert\omega_n\vert/{\bf q}^2\quad,
\label{eq:4.2a}
\end{equation}
with, 
\begin{equation}
u_2({\bf q}) = u_2^{(D-2)}\vert{\bf q}\vert^{D-2} + u_2^{(0)}{\bf q}^2\quad.
\label{eq:4.2b}
\end{equation}
\end{mathletters}%
Here $u_2^{(D-2)}$ and $u_2^{(0)}$ are finite numbers.
Note that for $D < 4$ (in particular, in the physical dimension $D=3$),
the first term in Eq.\ (\ref{eq:4.2b}) dominates the second one as 
${\bf q}\rightarrow 0$. $u_4$, in wavenumber space at zero frequency, is
schematically given by
\begin{equation}
u_4({\bf q}\rightarrow 0) = u_4^{(D-6)}\vert{\bf q}\vert^{D-6} 
                                + O(\vert{\bf q}\vert^{D-4})\quad,
\label{eq:4.3}
\end{equation}
i.e., $u_4$ diverges for $D < 6$. The singularities in the wavenumber
dependences of $u_2$ and $u_4$ mean that the field theory is non-local.
As mentioned above, their physical origin are
diffusive fermionic particle-hole excitations that were integrated out in
deriving Eq.\ (\ref{eq:4.1}).

Next we argue that in at least one well defined physical situation, it is
easy to uncover the effects of rare regions that are implicit in 
Eq.\ (\ref{eq:4.1}). 
The basic argument is that on length scales small compared to the elastic 
mean free path $\ell$, the field theory is effectively local. 
This implies that as long as the local moments or instantons decay on a 
scale $\lambda < \ell$, the techniques discussed in Sec.\ \ref{sec:II} 
can be used to include the effects of
these inhomogeneous saddle points on the final long-wavelength theory.
Further we will show that the quenched randomness that leads to these local
moments is implicitly contained already in the last term in 
Eq.\ ({\ref{eq:4.1}). To this end we first note, cf. Ref.\ \onlinecite{fm} 
and below, that the {\em exact} critical behavior 
near the ferromagnetic transition can be determined from 
Eq.\ (\ref{eq:4.1}). Second, we partially undo the replica trick in 
Eq.\ (\ref{eq:4.1}) by writing the logarithm of the
partition function, or the free energy, as Eq.\ (\ref{eq:2.15}) with $Z$ on 
the right hand side given by, 
\begin{mathletters}
\label{eqs:4.4}
\begin{equation}
Z = \int D[{\bf M}(x)]\ \exp \left( -S[{\bf M}(x),\delta t({\bf x})]
   \right)\quad,
\label{eq:4.4a}
\end{equation}
with, 
\begin{eqnarray}
S[{\bf M}(x),\delta t({\bf x})]&=&\frac{1}{2}\int dx_1\,dx_2\ \Gamma(x_1-x_2)
\nonumber\\
&&\hskip 70pt\times{\bf M}(x_1)\cdot{\bf M}(x_2)
\nonumber\\
&&\hskip -40pt +\int dx_1\,\ldots\,dx_4\ u_4(x_1,x_2,x_3,x_4)
\nonumber\\
&&\times({\bf M}(x_1)\cdot{\bf M}(x_2))
   \,({\bf M}(x_3)\cdot{\bf M}(x_4))\quad.
\nonumber\\
\label{eq:4.4b}
\end{eqnarray}
\end{mathletters}%
$\Gamma (x)$ is given by
\begin{equation}
\Gamma (x) = \Gamma_0(x) + \delta t({\bf x})\quad,
\label{eq:4.5}
\end{equation}
with $\Gamma_0$ from Eq.\ (\ref{eq:4.2a}). $\delta t(\mathbf{x)}$ 
in Eqs.\ (\ref{eqs:4.4}) is
a random function of position that is Gaussian distributed with the first
two moments given by Eqs.\ (\ref{eqs:2.2}). It trivally follows that 
Eqs.\ (\ref{eq:2.15}), (\ref{eq:4.3}) - (\ref{eq:4.5}), and (\ref{eqs:2.2}) 
are equivalent to Eq.\ (\ref{eq:4.1}).

The arguments given in Sec.\ \ref{sec:II} for the AFM case apply equally 
well to the action given by Eq.\ (\ref{eq:4.4b}). To make this more precise 
let us consider the non-localities. In terms of scaled variables, 
Eq.\ (\ref{eq:4.2b}) for $u_2(q)$ can be written to lowest order in the 
disorder as, 
\begin{equation}
u_2({\bf q}) = u_2^{(0)}\left[\left(\frac{{\bf q}}{k_F}\right)^2
   +\frac{c}{k_F\ell}\left(\frac{\vert{\bf q}\vert}{k_F}\right)^{D-2}\right]
     \quad, 
\label{eq:4.6}
\end{equation}
with $k_F$ the Fermi wavenumber, $\ell $ the elastic mean free path, and $c$
an interaction dependent constant that is at most of order unity. In
general the expansion implied by Eq.\ (\ref{eq:4.6}) assumes that 
$\vert{\bf q}\vert \ll k_F$.
In terms of length scales let $\lambda \sim 1/\vert{\bf q}\vert$ be 
the scale over which
the order parameter varies, and $\lambda_F\sim k_F^{-1}$ the Fermi
wavelength. Further, to be specific we consider the physical dimension $D=3$. 
The analytic, square gradient, term in Eq. (\ref{eq:4.6}) then 
dominates the second
term when $\lambda \ll \ell $. That is, the non-locality in 
Eq.\ (\ref{eq:4.6}) is
irrelevant when spatial scales shorter than a mean free path are considered.
For $\lambda _F \ll \lambda \ll \ell $ we then have, 
\begin{equation}
u_2({\bf q})\simeq u_2^{(0)}\left(\frac{\bf q}{k_F}\right)^2\quad. 
\label{eq:4.7}
\end{equation}
Note that for this argument to be valid we need 
$\lambda_F/\ell \approx 1/{k_F\ell}\ll 1,$ i.e., weak disorder 
is required. Similarly, $u_4$ in Eq.\ (\ref{eq:4.3}) can be replaced
by a constant when $\lambda_F \ll \lambda \ll \ell$. The net 
result is that when the local moments vary on a length scale smaller 
than $\ell$, they can be described by a local field theory analogous to 
the one discussed in Sec.\ \ref{sec:II} even though the long-wavelength 
theory is non-local. If we assume, as we did in the antiferromagnetic case
in the previous section, that the island distribution falls off
exponentially for large island sizes, this will always be true for
sufficiently small disorder.

With the above ideas and the techniques developed in Sec.\ {\ref{sec:II}, 
the final
long wavelength theory to descibe the ferromagnetic phase transition,
explicitly including the effects of rare regions, is determined by the
action, 
\begin{mathletters}
\label{eqs:4.8}
\begin{equation}
S_{\rm eff} = S_{\rm eff,1} + \delta S_{\rm eff},
\label{eq:4.8a}
\end{equation}
with $S_{\rm eff,1}$ given by Eq.\ (\ref{eq:4.1}) and
\begin{eqnarray}
\delta S_{\rm eff}&=&-w\sum_{\alpha}\int d{\bf x}\,d\tau\,
   d\tau^{\prime}\ ({\bf M}^{\alpha}(x,\tau)\cdot{\bf M}^{\alpha}(x,\tau))\,
\nonumber\\
 &&\hskip 30pt\times ({\bf M}^{\alpha}(x,\tau^{\prime})
      \cdot{\bf M}^{\alpha}({\bf x},\tau^{\prime}))\quad,
\label{eq:4.8b}
\end{eqnarray}
\end{mathletters}%
where $w$ is a finite constant.

Power counting immediately reveals that the coupling constant $w$, just
like $\Delta$, is an irrelevant operator with respect to the Gaussian 
fixed point discussed in Ref.\ \onlinecite{fm}. The rare regions therefore
do not change the critical behavior in this case. The physical reason for
this is the effective long-range interaction between the order parameter
fluctuations that is described by the $\vert{\bf q}\vert^{D-2}$-term in
Eq.\ (\ref{eq:4.2b}). This stabilizes the Gaussian fixed point by
suppressing all fluctuations, including the static disorder fluctuations
resonsible for the local moments.

\section{Discussion and Conclusion}
\label{sec:V}

In this section, we conclude by discussing the results obtained in the
previous sections.

\subsection{General considerations}
\label{subsec:V.A}

We begin our discussion by considering the physical underpinnings of 
some general aspects of our technical procedure. 

\subsubsection{Local moments and annealed disorder}
\label{subsubsec:V.A.1}

Let us first of
all give a simple physical interpretation of the $w$-term in the effective 
action, Eq.\ (\ref{eq:2.19}), which is the most important of the contributions
that reflect the existence of rare regions and local moments. Since the
local moments are self-generated by the electronic system, in response to
the potential created by the quenched disorder, they are an integral part
of the system and in equilibrium with all other degrees of freedom. In
our approximation, which takes into account only the static local moment
fluctuations, the effect of the rare regions therefore amounts to the
existence of static, annealed disorder. Indeed, a straightforward
generalization of Eq.\ (\ref{eq:2.10}) is to integrate over a manifold
of saddle points ${\bf\Phi}({\bf x})$, weighted with an appropriate 
distribution $P[{\bf\Phi}({\bf x})]$,
\begin{equation}
Z \hskip -2pt \approx \hskip -2pt \int D[{\bf\Phi}({\bf x})]\,
   P[{\bf\Phi}({\bf x})]\, \int_{<} D[\boldvarphi(x)]\
   e^{-\Delta S[\Phi({\bf x}),\boldvarphi(x),\delta t({\bf x})]},
\label{eq:5.1}
\end{equation}
which makes obvious the annealed-disorder character of the average over
the saddle points. The detailed result of the integration over the saddle
points will of course depend on the distribution $P$, which in turn
depends on the microscopic details of the disorder realization in the system.
However, any physically reasonable distribution will lead in particular to
a term in the effective action that has the structure of the $w$-term in 
Eq.\ (\ref{eq:2.19}). Since the saddle points are separated by large energy 
barriers in configuration space (see Sec.\ \ref{subsubsec:V.A.2} below), 
this term clearly cannot be
obtained by perturbatively expanding about the trivial homogeneous
saddle point as is done in the conventional theory. Thus, our method
approximately takes into account what one would call `non-perturbative'
effects in the usual approach.

It is important to note that the new term in the action, Eq.\ (\ref{eq:2.19}),
differs from the usual quantum fluctuation or $u$-term only in its time
structure. In the classical limit, therefore, $w$ just renormalizes $u$,
decreasing its bare value. This is indeed well known to be the only effect
of static annealed disorder in classical systems. In their analysis of
classical magnets, the authors of Ref.\ \onlinecite{Dotsenko} therefore
considered a more elaborate scheme for doing the sum over saddle points
in Eq.\ (\ref{eq:2.10}), or the integral in Eq.\ (\ref{eq:5.1}), that
leads to a term that breaks the replica symmetry. Our way of approximating
that integral can be considered as a zeroth order step in the approximation
scheme of Ref.\ \onlinecite{Dotsenko}. In the quantum case, the time structure
results in this zeroth step already giving a non-trivial result, and in this
sense quantum systems are more sensitive to rare region effects than
classical ones. The physical meaning of replica symmetry breaking in this
context is not quite clear. However, in the quantum case it is not necessary
to enter into this discussion. The AFM fixed point is unstable already under
the effects considered above, and no new fixed point exists. 
Considering replica
symmetry breaking in addition to our effect would not change this conclusion.
In the FM case, it turns out that the previously found Gaussian fixed point
is stable against replica symmetry breaking as well as against the quantum
effect, as we will discuss in more detail below.

\subsubsection{Energy barriers between saddle points}
\label{subsubsec:V.A.2}

A question that arises in connection with Eq.\ (\ref{eq:2.9}) or 
(\ref{eq:5.1}) is whether it is really true that there are large energy
barriers between the various saddle-point configurations, as our approximation
for the partition crucially depends on this assumption. Let us first consider
the case of the Ising model ($p=1$), for which we performed the explicit 
derivation in Sec.\ \ref{sec:II}. Suppose we have two saddle points 
that differ only by the sign of the order parameter on one particular 
island. In order to turn one of these spin configurations into the other, 
we need to flip all of the spins on that island. (For simplicity, we refer 
to the order parameter field as `spins'.) In order to do so, one must go 
through an intermediate state with a domain wall across the island. The 
energy of that domain wall can be estimated from the squared gradient
term in the free energy, integrated over the island, 
$J\int d{\bf x}\ (\nabla\phi({\bf x}))^2$, with $J$ the coupling between the
spins. The thickness of the domain wall is a microscopic length $a$,
and hence the energy of the domain wall, or the energy
barrier between the two saddle points, is proportional to 
$J\,L^{D-1}a/a^2 = J\,L^{D-1}/a$, with $L$ the linear size of the island. 
In the case of a continuous spin model ($p>1$) an analogous 
argument holds, except that now all length scales are
of order $L$.\cite{Grinstein} This leads to $J\,L^{D-2}$ for the
energy of a domain wall. For $D>2$, there
is thus only a quantitative difference between the Ising case and the
continuous spin case. 

In either model, the domain wall energy will have
to be multiplied by the number of islands by which two typical saddle
points differ. For the Ising case, let us consider $N$ islands, with
$2^N$ saddle points and $2^{N-1}(2^N-1)$ pairs of saddle points. The
probability distribution $\{p_N(n)\}$, for a pair of saddle points to 
have $n$ islands that are different is easily found to be
\begin{mathletters}
\label{eqs:5.2}
\begin{equation}
p_N(n) = \frac{1}{2^N-1}\,{N\choose n}\quad.
\label{eq:5.2a}
\end{equation}
For large $N$, this becomes a Gaussian distribution with mean $N/2$
and variance $\sqrt{N}/2$,
\begin{equation}
p_{N\rightarrow\infty}(n) = \frac{2}{\sqrt{2\pi N}}\,e^{-2(n-N/2)^2/N}\quad.
\label{eq:5.2b}
\end{equation}
\end{mathletters}%
One expects this to be true for the continuous spin case as
well, although the statistical analysis becomes much more involved in
that case. The miscroscopic energy of a domain wall thus gets multiplied
by a macroscopic number, leading to energy barriers between almost all
pairs of saddle points that go to infinity in the thermodynamic limit.
This justifies our approximation.

Finally, we note that our considerations maximize, and probably overestimate, 
the effects of local moments or disorder induced
instantons. The discussion above seems to imply an exponential number of
saddle point solutions that are unrelated by symmetries, with barriers
between them that approach infinity in the bulk limit.
This in turn implies an exponential number of thermodynamic states,
or a finite complexity. Such a proposition is
controversial in other contexts, e.g. for spin glasses. 
However, in our considerations we have
effectively neglected the interactions between the local moments. One
anticipates these interaction to correlate and weaken the rare regions.
Indeed, in Ref.\ \onlinecite{BhattFisher} it was argued that long-range 
interactions that arise
from the itinerant nature of the electrons quench most of the local
moments. If this happens in the systems we consider, then we
likely overestimate the number of distinct thermodynamic states. It is also
possible that our theory is valid only in an intermediate time region, and
that the interactions between the local moments must be
taken into account in the limit of asymptotically long times.

\subsubsection{Nature of the local-moment phase}
\label{subsubsec:V.A.3}

Another point we have not yet addressed is the physical nature of the
phase that is induced by the presence of the local moments. In order to
show that we are dealing with a Griffiths phase, 
let us consider the local moment contribution to the order
parameter susceptibility, $\chi_{\rm LM}$. Let us adopt a ferromagnetic
language for simplicity, and denote the magnetic moment on the island
number $i$ by $M_i$. Then we have
\begin{eqnarray}
\chi_{\rm LM} = \biggl\{\frac{1}{\sum_i V_i} \int_{0}^{1/T} d\tau\sum_{ij}\left(
   \langle M_i(\tau)\,M_j(0)\rangle\right.
\nonumber\\
 - \left.\langle M_i\rangle\,\langle M_j\rangle \right)
       \biggr\}_{\delta t}\quad,
\label{eq:5.3}
\end{eqnarray}
where $\langle\ldots\rangle$ denotes a thermodynamic average. Since there
is no overall magnetization, $\sum_i \langle M_i\rangle = 0$, and in our 
saddle-point approximation the island magnetization is static. This yields
\begin{equation}
\chi_{\rm LM} = \biggl\{\frac{1}{\sum_i V_i}\,\frac{1}{T}\,
   \sum_{ij}\langle M_i\,M_j\rangle\biggr\}_{\delta t} 
   = \frac{\rm const}{T}\quad,
\label{eq:5.4}
\end{equation}
where the constant is given by 
$\{\sum_i \langle M_i^2\rangle/\sum_i V_i\}_{\delta t}$.
We see that the order parameter susceptibility 
diverges for $T\rightarrow 0$ whenever there are islands, and in our
simple saddle-point approximation the divergence takes the form of a
Curie law. Our saddle point thus really describes a Griffiths phase.

\subsubsection{Finiteness of the free energy}
\label{subsubsec:V.A.4}

As a final general point, let us again consider the effective action,
Eq.\ (\ref{eq:2.19}), which determines the free energy. Since the
$w$-term and the $u$-term have the same structure except for an extra
time integral in the former, it seems as if the $w$-term contributes a
term to the free energy that diverges as the temperature goes to zero.
One has to keep in mind, however, that Eq.\ (\ref{eq:2.19}) represents
a Landau expansion that has been truncated at $O(\boldvarphi^4)$. It
is easy to see that higher order terms in the Landau expansion lead to
even more strongly divergent contributions to the free energy, see
Eq.\ (\ref{eq:3.4}). This simply means that the loop expansion for
the free energy of a quantum system with static annealed disorder is
singular, and a resummation to all orders would be necessary to obtain
a finite result. From a RG point of view, which holds that the higher
order terms in the Landau expansion are irrelevant, the solution of this
paradox lies in the fact that, if a fixed point existed, it would be 
${\bar w}$ that has a finite fixed-point value, not $w$. Since 
$w = {\bar w}\,T$ (for the physical case $\epsilon_{\tau}=1$), this 
ensures that the fixed-point Hamiltonian has a finite free energy.

\subsection{Results for the AFM case}
\label{subsec:V.B}

As we have shown in Sec.\ \ref{sec:III}, and reiterated above, taking into
account the rare regions in the AFM case destroys the stability of
the fixed point found in Ref.\ \onlinecite{afm}, and one finds
runaway flow in all of the physically accessible parameter space.
Three possible interpretations of this result are,
(i) there is no transition to a state with long-range order,
(ii) there is a transition, but the corresponding fixed point is inaccessible
by perturbative RG techniques, or (iii) there is a fluctuation-induced first
order transition (which causes the runaway flow). The last conjecture
can be checked by calculating the free energy
to one-loop order and then explicitly verifying whether it has
a double minimum structure as a function of the order parameter.
We have performed such a calculation,\cite{rnthesis} and found that this
is not the case. This rules out scenario (iii).

On the basis of our results, we cannot decide between scenarios (i) and (ii).
Scenario (i) would imply that arbitrarily weak disorder necessarily destroys
quantum AFM long-range order. This is an unlikely proposition, but it cannot
be ruled out at present. The alternative is scenario (ii),
i.e. the existence of a non-perturbative fixed point.
The nature of such a fixed point, if it exists, is a priori unclear.
The analogies with $1$-$D$ systems mentioned in the Introduction, as well
as Ref.\ \onlinecite{APY2d}, suggest that an unconventional infinite
disorder fixed point with
activated scaling is a possible interpretation of the runaway flow. However,
there also could be a conventional fixed point that is not accessible by
our methods. In this context it is interesting to note that the case
$p>4$ discussed in Sec.\ \ref{subsec:III.C} provides an example of a stable
conventional fixed point that describes a transition with power-law
critical behavior in the presence of rare regions.

Let us also come back to the fact that in Sec.\ \ref{sec:III} we found
a stable fixed point (No. 8 in Table \ref{tab:1}) with ${\bar w}^* \neq 0$.
As was pointed out in Sec.\ \ref{subsec:III.C}, for generic realizations
of the disorder, which lead to a positive bare value of $w$, this fixed 
point is unphysical since it has ${\bar w}^* < 0$. However, mathematically
one can have $w<0$ for certain choices of the distribution $P[\{\sigma_i^a\}]$
in Sec.\ \ref{sec:II} that are more general than Eq.\ (\ref{eq:2.7c}). 
This leaves open the possibility that at least in
some systems there is a stable, conventional critical fixed point that is
accessible with our method. We note that this fixed point is stable against
replica symmetry breaking, see Appendix \ref{app:C}. This is in contrast to
the case of classical magnets,\cite{Dotsenko} where all fixed points are
unstable against replica symmetry breaking, and reminiscent of the result
of Read et al.\cite{Readetal} on quantum spin glasses, where the quantum
model was also found to be more stable against replica symmetry breaking
than its classical counterpart. The technical reason for this enhanced
stability is very similar to the point discussed at the end of Sec.\ 
\ref{subsec:V.A}, namely that the parameter that would induce replica
symmetry breaking appears as $T$ times a finite fixed point value, and
hence vanishes at the quantum critical point.

\subsection{Results for the FM case}
\label{subsec:V.C}

For itinerant quantum ferromagnets, we have found that the rare regions do not
affect our previous results.\cite{fm} The physical reason for this is the
long-range interactions between the spin fluctuations in these systems.
They are induced by soft modes in the itinerant electron system and stabilize
the Gaussian critical behavior against fluctuations, including the static
disorder fluctuations that lead to local-moment formation. A crucial
point for our conclusion is the survival of these long-range interactions
in the presence of local moments, so it is worth discussing this in some
detail.

The derivation of the long-range interaction\cite{fm} shows that its origin
is soft spin-triplet particle-hole excitations in the electron system.
An obvious question is whether local moments act effectively as magnetic
impurities that give these soft modes a mass. If this were the case, then
the singular wavenumber dependences
$\vert{\bf q}\vert^{d-2}$ and $\vert\omega_n\vert/{\bf q}^2$ in the Gaussian 
vertex, Eq.\ (\ref{eq:4.2a}), would be cut off and the ferromagnetic
effective action would have the same structure as the antiferromagnetic
one. The answer to this question is not obvious since the local moments
are self-generated, and thinking about them as analogous to externally
introduced magnetic moments can be misleading. This is underscored by
the fact that the rare-regions/local-moment physics enters the theory
in the form of {\em annealed} disorder, as we have seen in 
Sec.\ \ref{subsec:V.A} above. In our derivation of the effective action,
Sec.\ \ref{sec:IV}, the wavenumber singularities are not cut off.
The physics behind this
is that both singularities are consequences of spin diffusion, which in
turn is a consequence of the spin conservation law. The rare regions
ultimately derive from a spin-independent disorder potential, which
clearly cannot destroy spin conservation. The long-range interactions
between the spin fluctutations are therefore still present in the bare 
effective action, and hence the Gaussian fixed point is stable in our
tree-level analysis. We note, however, that at present we cannot rule
out the possibility that loop corrections might lead to qualitatively
new terms in the action. If such new terms included a RG-generated spin 
dependent potential, then this might change our conclusions. This would
not necessarily violate the spin conservation argument given above, since
an effective spin-dependent potential acting only on the itinerant electrons,
which are not taking part in the local moment formation, could change the
critical behavior while preserving spin conservation for the system as a
whole. 

A more detailed investigation of this point is not feasible within the
existing framework of the ferromagnetic theory.\cite{fm} This is because
in the existing theory all degrees of freedom other than the order parameter,
including various soft modes, have been integrated out. This leads to a 
non-local field theory which is unsuitable for an explicit loop expansion.
A remedy would be to derive an effective theory that keeps all soft modes
explicitly and treats them on equal footing, leading to well-behaved
vertices that allow for explicit calculations. This project is left for
future work. We also note that at such a level of the
analysis one should also include effects due to interactions
between the rare regions, which we have mostly neglected. Such interactions
are known to weaken the effects of the rare region,\cite{BhattFisher}
but in general it is not known by how much.

\subsection{Summary, and Outlook}
\label{subsec:V.D}

In summary, we have studied the effects of rare disorder fluctuations, and
the resulting local moments, on itinerant ferromagnets and antiferromagnets.
Technically, this has been achieved by considering non-trivial saddle-point
solutions before performing the disorder average. A perturbative RG analysis
of the resulting effective field theory incorporates effects that would
require non-perturbative methods within a more standard procedure.
In the ferromagnetic case we have found that, at least within our level
of analysis, the previously found quantum critical behavior\cite{fm} is stable
with respect to local moment physics. The reason is an effective long-range
interaction between the spin fluctuations that strongly suppresses
fluctuations, stabilizing a Gaussian critical fixed point. In the 
antiferromagnetic case, however, we have found that the local moments
destroy the previously found critical fixed point.\cite{afm} 
To one-loop order, and for order parameter dimensionalities less than $4$, no
new fixed point exists and one finds runaway flow in all of physical
parameter space. This may indicate either the absence of long-range order,
or a transition that is not perturbatively accessible within our theory.

An important technical conclusion is that for quantum phase transitions, and
within the framework of a replicated theory, rare regions can have a
qualitative effect already at the level of a replica-symmetric theory,
in contrast to the case of classical magnets.\cite{Dotsenko} 
The ferromagnetic fixed point, which was found to be stable against the 
replica-symmetric quantum effects induced by the rare regions, is also 
stable against replica-symmetry breaking.

We have concentrated on the role of fluctuations about a non-trivial, but
fairly crudely constructed, saddle-point solution of the field theory. It
would also be interesting to study a somewhat more sophisticated saddle-point
theory in more detail, and to determine the detailed properties of the
Griffiths phase in such an approximation.

Finally, we mention that our methods are not specific to magnets, and can
be applied to other quantum phase transitions as well. For instance, it
is believed that for a complete understanding of the properties of doped
semiconductors, and of the metal-insulator transitions observed in such
systems, it is necessary to consider the effects of local 
moments.\cite{BhattLee,BhattFisher,R} This can be studied with the methods
developed in this paper.

\acknowledgments
We are indebted to Ferdinand Evers for a collaboration in the early
stages of this work. We also gratefully acknowledge helpful discussions 
with John Toner. This work was supported by the NSF under 
grant Nos. DMR-95-10185, DMR-96-32978, and DMR-98-70597, and by the DFG under 
grant No. SFB 393/C2.

\appendix
\section{A one-dimensional saddle-point equation}
\label{app:A}

In this Appendix we discuss the saddle-point equation, Eq.\ (\ref{eq:2.5}),
for a particular realization of the disorder potential $\delta t({\bf x})$.
In particular, we aim to show that the existence of a non-zero solution
requires the width and depth of the potential well to be above a threshold,
and that the non-zero solution lowers the free energy with respect to the
zero one.

We first consider the one-dimensional counterpart of
Eq.\ (\ref{eq:2.5}),
\begin{equation}
\left(t_0 + \delta t(x) - \partial_x^2\right)\,\phi(x) + g\phi^3(x) = 0\quad,
\label{eq:A.1}
\end{equation}
with a simple square well potential,
\begin{equation}
\delta t(x) = \cases{-V_0 & for $0 \le x <a$\cr
                  0 & elsewhere\cr} \quad.
\label{eq:A.2}
\end{equation}
Standard methods lead to a solution inside the well ($0\leq x < a$)
\begin{mathletters}
\label{eqs:A.3}
\begin{equation}
\phi_{\rm in} (x) = \sqrt{\frac{v_1}{g}}\,
       \frac{{\rm cn}\sqrt{v_2\,x/2}}{{\rm dn}\sqrt{v_2\,x/2}} \quad,
\label{eq:A.3a}
\end{equation}
where ${\rm cn}$ an ${\rm dn}$ are elliptic functions,
and a solution outside of the well,
\begin{equation}
\phi_{\rm out}(x) = \sqrt {\frac {t_0}{g}}\,\frac {2\,\sqrt{2}\,c\,e^{-\,
    \sqrt {t_0}\,x}} {c^2\, e^{-2\, \sqrt {t_0}\, x} - 1}
\label{eq:A.3b}
\end{equation}
\end{mathletters}%
Here
\begin{mathletters}
\label{eqs:A.4}
\begin{equation}
v_{2,1} = \frac {\alpha \pm \sqrt{{\alpha}^2 - 4\, \beta}}{2}\quad,
\label{eq:A.4a}
\end{equation}
with
\begin{equation}
\alpha = 2(t_0-V_0) \quad.
\label{eq:A.4b}
\end{equation}
\end{mathletters}%
$c$ and $\beta$ are constants of integration that are determined by the
requirement that the solution and its derivative be continuous at $x=a$.
Other solutions exist, but the one given is the only one that satisfies
physical boundary conditions. Furthermore, the physical solution exists
only for
\begin{equation}
0 < \beta < \alpha^2/4 \quad.
\label{eq:A.5}
\end{equation}

To demonstrate the existence of a threshold, we expand the above solution
for small values of $v_2a$. To leading order in this small parameter,
we obtain for the constants of integration
\begin{mathletters}
\label{eqs:A.6}
\begin{equation}
c = \frac {1}{4}\, \sqrt {\frac {2}{t_0}}\, \sqrt{v_1}\,e^{\sqrt{t_0}\,a}\quad,
\label{eq:A.6a}
\end{equation}
and
\begin{equation}
\beta = \frac {{\alpha}^2}{4} - \frac {t_0}{2\, a^2} \quad.
\label{eq:A.6b}
\end{equation}
\end{mathletters}%
From the condition for the existence of the physical solution,
Eq.\ (\ref{eq:A.5}), we see that $\alpha^2 a^2 > 2t_0$ is a necessary and
sufficient condition for the solution to exist, which is the desired
threshold property.

The free energy in saddle-point approximation is simply given by
the saddle-point action. It is physically plausible that the non-homogeneous
solution constructed above leads to a negative free energy, and is thus
energetically favorable compared to the zero solution. We have ascertained
this numerically for a large variety of well parameter values, and have found
the inhomogeneous solution to lead to a negative free energy whenever it
exists.
\begin{figure*}[t]
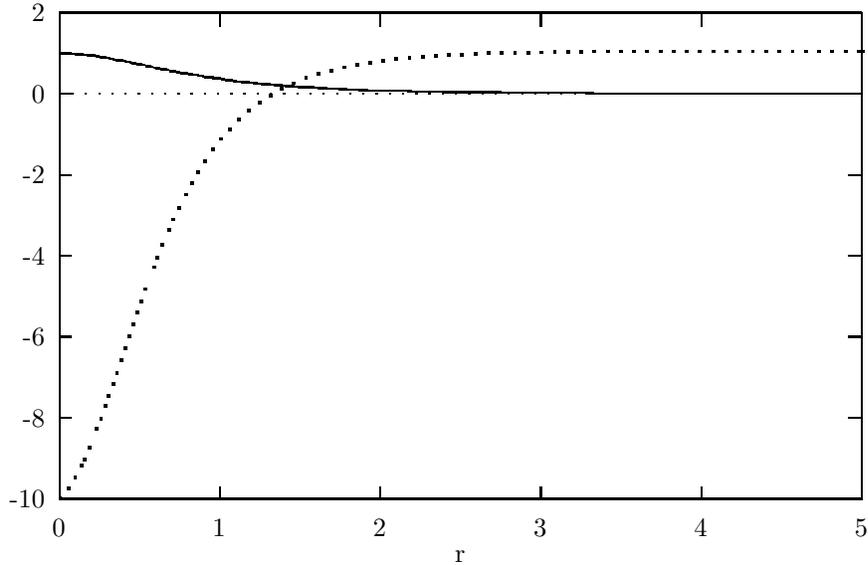

\centerline{\input fig_2.tex}
\caption{$t(r)$ (dotted line) and $\phi (r)$ (solid line) for $a=1, b=1, c=1$.}
\label{fig:2}
\end{figure*}

To solve the three-dimensional saddle-point equation, Eq.\ (\ref{eq:2.5}),
is much harder. For sperically symmetric wells, $t_0+\delta t({\bf x})=t(r)$,
with $r=\vert{\bf x}\vert$, analytic solutions can still
be found in closed form for special forms of the potential. By scaling
$\phi$ and $t$, the equation can be written
\begin{equation}
\nabla^2 + t(r)\,\phi(r) + \phi^3(r) = 0 \quad.
\label{eq:A.7}
\end{equation}
It is easy to show that for
\begin{mathletters}
\label{eqs:A.8}
\begin{eqnarray}
t(r) &=&  \frac {-2b}{(1 + b\, r)} + b^2 + \frac {4\, b\, c\, r}{(1 + c\, r^2)}
 -  \frac {2\, c}{(1 + c\, r^2)}
\nonumber\\
 &-& \frac {4\, b\, c\, r}{(1+c\, r^2)\, (1+b\, r)} + \frac {8\, c^2\, r^2}
          {(1+c\, r^2)^2} - \frac {2\, b^2}{(1+b\, r)}
\nonumber\\
 &-& \frac {4\, c}{(1+b\, r^2)} - \frac {a^2\, (1+b\, r)^2\, e^{-2\, b\, r}}
           {(1+c\, r^2)^2} \quad,
\label{eq:A.8a}
\end{eqnarray}
the physical solution is given by
\begin{equation}
\phi (r) = \frac {a\,(1 + b\,r)\, \exp ({-b\,r})}{(1 + c\,r^2)}\quad.
\label{eq:A.8b}
\end{equation}
\end{mathletters}%
Here $a$, $b$, and $c$ are parameters that determine
the shape of the well.  In this case, the physical solution exists for all
real values of the three parameters, but the form of the potential is such
that the volume of the well cannot be smaller than some minimum value.
This is the three-dimensional analog of
the threshold behavior demonstrated above for the one-dimensional case.
We have also solved the ODE, Eq.\ (\ref{eq:A.7}), numerically for more
general potential wells, and have found the same type of threshold
behavior.

As in the $1$-$D$ case, physical arguments suggest, and numerical 
integration confirms,
that the inhomogeneous solution leads to a lower free energy than the
homogeneous one whenever the former exists.
In Fig.\ \ref{fig:2} we show the solution and the corresponding potential
well, Eqs.\ (\ref{eqs:A.8}), as a representative example of a locally ordered
region in a $3$-$D$ system.

\section{Island size distributions with a power-law tail}
\label{app:B}

In Secs.\ \ref{sec:III} and \ref{sec:IV} we have assumed that the island
correlation functions $D^{(m)}_{\rm isl}$ and $C^{(m)}_{\rm isl}$
are short-range correlated, i.e. have a scale dimension of $(m-1)D$. Here
we briefly discuss the extent to which one can relax this condition 
without changing our results.

Suppose that the island-size distribution is power-law correlated, leading
to scale dimensions of the above correlation functions that are given by
$(m-1)(D-\alpha)$ with $\alpha>0$. Let us consider the FM case first.
The least irrelevant term, viz. Eq.\ (\ref{eq:4.8b}), remains
irrelevant with respect to the Gaussian fixed point of Ref.\ \onlinecite{fm}
as long as $\alpha < 4-D$ (for $2<D<4$). The $D$-dependence of this result 
reflects the fact that for $D\geq 4$ the effective interaction ceases to be 
long-ranged, and an ordinary mean-field fixed point is stable. All higher
order terms in the action are less relevant than the $w$-term.

In the AFM case, the $w$-term is relevant with respect to the conventional
fixed point even for $\alpha = 0$. By power counting, we find the condition
that none of the higher order terms become relevant as well, viz.
$\alpha < D-3$ for $D$ close to $4$. Here the $D$-dependence reflects the
fact that the coupling constant $w_6$ is marginal in $D=3$ even for
$\alpha = 0$, see Sec.\ \ref{subsec:III.A}.

\section{Stability under replica symmetry breaking}
\label{app:C}

In this Appendix we briefly consider the effects of replica symmetry
breaking (RSB). A generalization of our action, Eq.\ (\ref{eq:2.19}),
analogous to Ref.\ \onlinecite{Dotsenko} that allows for RSB is
\begin{eqnarray}
S_{\rm eff}[\boldvarphi^\alpha(x)] &=&
  \frac{1}{2} \sum_\alpha\,\int dx\,dy\,\boldvarphi^\alpha(x)\cdot\,
     \Gamma_0(x-y)\,\boldvarphi^\alpha(y)\
\nonumber\\
&&\hskip -40pt + u\sum_\alpha \int d{\bf x}\, d\tau
     \left(\boldvarphi^\alpha({\bf x},\tau)\cdot
          (\boldvarphi^\alpha({\bf x},\tau)\right)^2
\nonumber\\
&&\hskip -40pt -\sum_{\alpha,\beta}(\Delta + w_{\alpha\beta})\int d{\bf x}\,
   d\tau\,d\tau^{\prime}
\nonumber\\
&&\hskip -16pt \times (\boldvarphi^\alpha({\bf x},\tau)\cdot
                     \boldvarphi^\alpha ({\bf x}, \tau))\,
         (\boldvarphi^\beta ({\bf x},\tau^{\prime})\cdot
                        (\boldvarphi^\beta ({\bf x},\tau^{\prime}))\,.
\nonumber\\
\label{eq:C.1}
\end{eqnarray}
In Sec.\ \ref{sec:II} we had $w_{\alpha\beta}\equiv w$, which resulted in
a replica symmetric theory. Now we allow for 1-step RSB in Parisi's
hierarchical scheme,\cite{Parisi} where $w_{\alpha\beta}$ in the replica
limit is parameterized by means of a step function with a parameter $x_0$,
\begin{equation}
w(x) = \cases{w & for $0\leq x \leq 1$\cr
              w_1 & for $x_0 < x \leq 1$\cr}\quad.
\label{eq:C.2}
\end{equation}
Defining ${\bar w} = w\,T^{\epsilon_{\tau}}$ as before, and
${\tilde w} = w_1,T^{\epsilon_{\tau}}$, we obtain the 1-loop flow equations
\begin{mathletters}
\label{eqs:C.3}
\begin{equation}
\frac{du}{dl} = (\epsilon - 2\epsilon_{\tau})\,u
                              - 4\,(p+8)\,u^2 + 48\,u\,\Delta\quad,
\label{eq:C.3a}
\end{equation}
\begin{eqnarray}
\frac{d\Delta}{dl} &=& \epsilon\,\Delta + 32\,{\Delta}^2
             - 8\,(p+2)\,u\,\Delta + 8p\,\Delta\,{\bar w}
\nonumber\\
&-& 8p\,x_0\,\Delta\,{\tilde w} + 8p\,\Delta\,{\tilde w}\quad,
\label{eq:C.3b}
\end{eqnarray}
\begin{eqnarray}
\frac{d{\bar w}}{dl} &=& (\epsilon - 2\,\epsilon_{\tau})\,{\bar w} 
   + 4\,p\,{\bar w}^2 -8\,(p+2)\,u\,{\bar w} + 48\,\Delta\,{\bar w}
\nonumber\\
&&- 4p\,(1-x_0)\,{\tilde w}^2 \quad,
\label{eq:C.3c}
\end{eqnarray}
\begin{eqnarray}
\frac{d{\tilde w}}{dl} &=& (\epsilon - 2\, \epsilon_{\tau})\,{\tilde w} 
   + 48\,\Delta\,{\tilde w} - 8\,(p+2)\,u\,{\tilde w}
\nonumber\\
  &+& 8p\,{\bar w}\,{\tilde w} + 8p\,(1-x_0)\,{\tilde w}^2 \quad.
\label{eq:C.3d}
\end{eqnarray}
\end{mathletters}%
The replica symmetric case is recovered by putting $x_0 = 1$.
We now perform a linear stability analysis of fixed point No. 8, with 
fixed-point values of $u$, $\Delta$ and $w$ as given in Table \ref{tab:1}, and
${\tilde w}^* = 0$. The first three eigenvalues are as shown in
Table \ref{tab:1}, and the fourth one is 
$\lambda_{\tilde w} = \lambda_w = (p-4)(\epsilon + 4\epsilon_{\tau})/2(10-p)$.
Fixed point No. 8 is therefore stable against 1-step RSB. Although the fixed
point is unphysical for generic realizations of the disorder, as discussed
in Sec.\ \ref{sec:III}, this is an interesting contrast to the classical
case,\cite{Dotsenko} where all fixed points are unstable against successive
terms in the hierarchical RSB scheme.

\end{document}